PREPRINT vs 1

# Strain distribution in GaN/AlN superlattices grown on AlN/sapphire templates: comparison of X-ray diffraction and photoluminescence studies

*Aleksandra Wierzbicka[a], Agata Kaminska [a,b,c], Kamil Sobczak[d], Dawid Jankowski[e], Kamil Koronski[a], Pawel Strak[c], Marta Sobanska[a], Zbigniew R. Zytkiewicz[a]*

[a] Institute of Physics, Polish Academy of Sciences, Al. Lotników 32/46, 02-668 Warsaw, Poland

[b] Faculty of Mathematics and Natural Sciences, School of Exact Sciences, Cardinal Stefan Wyszynski University, Dewajtis 5, 01-815 Warsaw, Poland

[c] Institute of High Pressure Physics, Polish Academy of Sciences, Sokolowska 29/37, 01-142 Warsaw, Poland

[d] Faculty of Chemistry, Biological and Chemical Research Centre, University of Warsaw, Żwirki i Wigury 101, 02-089 Warsaw, Poland

[e] Institute of Physics, Faculty of Physics, Astronomy and Informatics, Nicolaus Copernicus University in Toruń, ul. Grudziądzka 5, 87-100 Toruń, Poland.

*Corresponding author: wierzbicka@ifpan.edu.pl*

KEYWORDS: superlattices, X-ray diffraction, Multi Quantum Wells, nitrides, strain relation

**ABSTRACT**

Series of GaN/AlN superlattices (SLs) with various periods and the same thicknesses of GaN quantum wells and AlN barriers have been investigated. X-ray diffraction, photoluminescence (PL) and transmission electron microscopy (TEM) techniques were used to study the influence of thickness of AlN and GaN sublayers on strain distribution in GaN/AlN SL structures. Detailed X-ray diffraction measurements demonstrate that the strain occurring in SLs generally decreases with an increase of well/barrier thickness. Fitting of X-ray diffraction curves allowed determining the real thicknesses of the GaN wells and AlN barriers. Since blurring of the interfaces causes deviation of calculated data from experimental results the quality of the interfaces has been evaluated as well and compared with results of TEM measurements. For the samples with thinner wells/barriers the presence of pin-holes and threading dislocations has been observed in TEM measurements. The best quality of interfaces has been found for the sample with a well/barrier thickness of 3 nm. Finally, PL spectra showed that due to Quantum-Confined Stark Effect the PL peak energies of the SLs decreased with increasing the width of the GaN quantum wells and AlN barriers. The effect is well modelled by ab initio calculations based on the density functional theory applied for tetragonally strained structures of the same geometry using a full tensorial representation of the strain in the SLs.

1. INTRODUCTION

III-nitride semiconductors are of considerable interest because their direct bandgap (Eg) covers wide spectral range (from $E_g(InN)$ = 0.65 eV at room temperature [1] through $E_g(GaN)$ = 3.43 eV [1, 2] to $E_g(AlN)$ = 6.03 eV [3, 4]). Materials based on group III nitride semiconductors are commonly used in optoelectronic devices [6], AlGaN-based Light Emitting Diodes (LEDs) [7], photodetectors [8] and laser diodes (LDs) [9]. The strain relation that occurs in AlGaN-GaN structures leads to generation of lattice defects and consequently to degradation of devices based on these materials. Therefore the fundamental examinations are essential for controlling the strain, and thus avoiding multiplication of defects. Much attention is concentrated on the structures grown on low cost substrates as sapphire or Si. But the problem of large lattice mismatch between GaN and $Al_2O_3$ (13.8 % [9]), AlN and $Al_2O_3$ (13.4 % [10]), GaN and Si (16.9 % [11]) or AlN and Si (19% [12]) remains. AlN interlayers between sapphire and GaN are often used to reduce the lattice mismatch. Therefore AlN/sapphire templates are often used as a substrate to grow SLs due to smaller lattice mismatch and higher thermal conductivity than GaN [11, 12, 13].

In our work, we use AlN/sapphire templates to grow series of SLs with various periods and the same thicknesses of GaN wells and AlN barriers. Fundamental structural analysis of the samples by XRD, TEM and PL techniques supported by ab initio calculations were applied to create a model of strain distribution in such structures.

2. SAMPLES AND EXPERIMENTAL METHODS

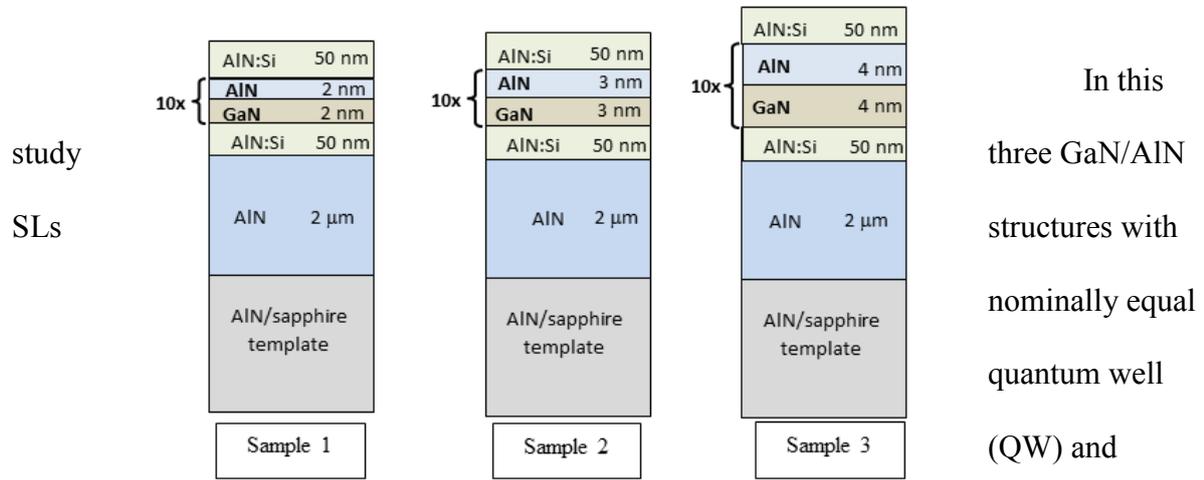

In this study three GaN/AlN SLs structures with nominally equal quantum well (QW) and quantum barrier (QB) thicknesses of 2, 3 or 4 nm were used. The structures were grown by plasma-assisted molecular beam epitaxy (PAMBE) on c-polar AlN/sapphire templates. The number of QW/QB periods was equal to 10. All three MQWs were capped with 50 nm of AlN layer doped with silicon to a concentration of $2 \times 10^{19} cm^{-3}$. The schematic drawing of the samples is presented in Figure 1.

**Figure 1**. Schematic cross-sections of the AlN/GaN MQWs structures studied in this work (not in scale).

The samples were systematically studied by X-ray diffraction (XRD) with the use of a laboratory Panalytical X'Pert PRO MRD diffractometer equipped with X-ray lamp with a wavelength of CuKα1, hybrid two-bounce Ge (220) monochromator and Soller slits (normal angular resolution) or triple-bounce Ge(220) analyzer (high angular resolution) in front of the Pixcel detector [16]. XRD measurements allow to analyze crystallographic orientation and to determine values of lattice parameters as well as strain relation in MQWs. Then, the samples were studied by Transmission Electron Microscopy (TEM) using TITAN CUBED 80 - 300 microscope operated at 300 kV, equipped with Cs-corrector and high-angle annular dark-field (HAADF) detector. Cross-sectional TEM specimens were prepared by a standard method of mechanical pre-thinning followed by Ar ion milling [17].

To investigate the optical properties of the samples the photoluminescence (PL) spectra were measured using as an excitation source deep-UV laser lines (275.4 or 300.2 nm) of the continuous wave (cw) Coherent Innova 400 Ar-ion laser, or a 325-nm line of a cw He-Cd laser. The excitation wavelength was chosen depending on the QW thickness, and hence the transition energies of the structure under study. The samples were mounted inside a closed-circle helium refriger ator. The spectra were dispersed by a Horiba Jobin-Yvon FHR 1000 monochromator, and the signal was detected by a liquid nitrogen-cooled charge-coupled device camera. More details about this type of investigation can be found elsewhere [18].

Finally, ab initio calculations were performed. Based on the density functional theory the strain relations in the MQWs were calculated [19], [20].

## 3. RESULTS AND DISSCUSION

First, the XRD measurements were done in the low-angle resolution mode. The θ/2θ scans (not shown here) show the [0001] orientation of the AlN/sapphire, AlN and AlN/GaN MQWs. Next, high-resolution XRD measurements were performed. Figure 2 shows experimental 2θ/ω scans of 0002 AlN/GaN reflection (black curves). The signals from 0006 sapphire reflection, from 0002 AlN substrate/buffer layer and 0, the 1st, 2nd (for Sample 1), 3rd (for Sample 3), and 4th (for Sample 2) order of MQWs reflections can be distinguished. Diffraction peaks were simulated using X'Pert epitaxy program, which is based on the dynamical theory of X'ray diffraction for distorted crystals [21], [22]. The scattering from periodic structures like MQWs is usually modelled with dynamical diffraction theory. Fewster considered also the conditions where the kinematical theory of XRD can be adopted [23]. In our case, however, the experimental results seems to be in good agreement with the dynamical theory of XRD, so it was used in this work. The simulations included the usage of hybrid 2-bounce monochromator (composed of an X-ray mirror and Ge (220) 2-bounce crystal). The peak position of 0006 sapphire substrate was determined and then the thicknesses and strain relations parameters were fitted. The simulation results are shown in Figure 2 as the red curves and are in good agreement with the experimental ones. However, the experimental and simulated curves differ in intensities. It is caused by the interface blurring and pinholes in our structures as will be discussed in the next part of this section when presenting TEM images of the samples.

The relative intensities of the XRD peaks coming from MQW structures are not the same for reversed order of satellite peaks. This is explained in detail in [24]. Note that the relative intensities of the satellite peaks in SL are determined by the envelope function which depends mainly on R – the well/barrier thickness ratio ($R = t_w/t_b$ where $t_w$ is the thickness of the well and $t_b$ is the thickness of the barrier). This function is proportional to the square of the structure factor of the multilayer. Thus, minima in the envelope function produce missing superlattice peaks, which are observed in the high-angle diffraction data collected over a wider range. The positions of the missing SL peaks are related to R. Although the small thickness fringes will be smeared out, the relative intensities of the SL peaks are preserved [24], [25], [26], [27].

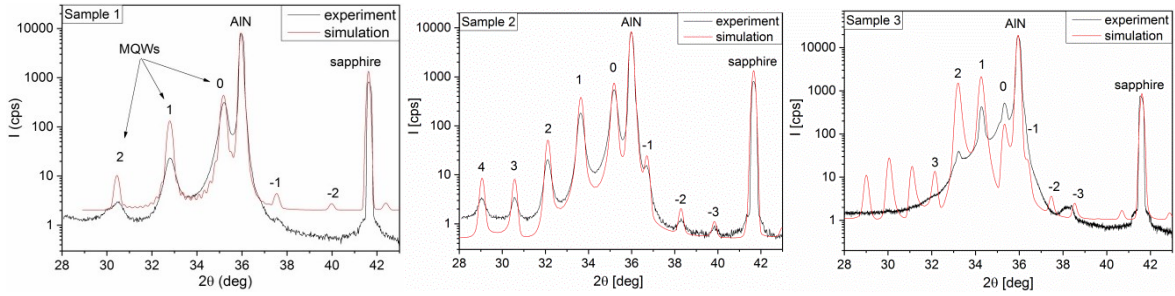

**Figure 2.** Experimental (black curves) and simulated (red curves) 2θ/ω diffraction curves of 00.2 $Cu_{K\alpha 1}$ reflections from the AlN/GaN MQWs structures.

The individual well-barrier and barrier-well thicknesses can be extracted from XRD scans also. The results obtained from XRD simulations are collected in Table 1. They are compared with sublayer thicknesses determined by high-resolution transmission electron microscopy (HTREM).

| Sample | Sample 1 | | Sample 2 | | Sample 3 | |
|---|---|---|---|---|---|---|
| | Thickness (nm) | | Thickness (nm) | | Thickness (nm) | |
| Layer | XRD | HRTEM | XRD | HRTEM | XRD | HRTEM |
| AlN | 1.9 | 1.9 | 3 | 2.8 | 2.3 | 2.5 |
| GaN | 2 | 1.9 | 3 | 2.9 | 6.3 | 6.9 |
| Total MQW | 39 | 38 | 60 | 57 | 86 | 94 |
| Lattice constants | $c_{MQW}$ = 5.0957 Å $a_{MQW}$ = 3.1209 Å | | $c_{MQW}$ = 5.0950 Å $a_{MQW}$ = 3.1275 Å | | $c_{MQW}$ = 5.0744 Å $a_{MQW}$ = 3.1457 Å | |

**Table 1.** Thicknesses of GaN wells and AlN barriers in MQWs obtained from XRD simulations and HRTEM measurements. Average lattice parameters of superlattices were calculated from HRXRD maps.

To determine the strain relations occurring in the superlattices the high-resolution XRD maps were registered. To avoid potential errors in determination of lattice parameters caused by tilt of lattice planes, we used an approach well described by Krysko et. al [28]. Briefly, the asymmetrical -1-124 reflections of AlN/GaN MQWs structures in six equivalent directions were measured. Then we chose the sample orientation in which the miscut direction was perpendicular to the diffraction plane. The interplanar distances of asymmetric planes measured in such configuration of the sample are not influenced by the tilt of the lattice planes. Figure 3 shows asymmetrical -1-124 reflections of AlN/GaN MQWs structures. As seen, the mutual positions of MQWs peaks with respect to AlN substrate are not on a vertical line in $Q_z$ ($Q_x$) coordinates demonstrating a partial relaxation of multi quantum wells relative to the substrate ($a_{SLs} \neq a_{AlN}$). Moreover, the arrangement of the MQWs peaks on HRXRD maps indicates the changes in *c*-lattice parameters as well. All values of the average MQWs lattice constants are shown in Table 1. Using them the relaxation degree of MQWs was calculated as equal to 30%, 27% and 30% for Sample 1, Sample 2 and Sample 3, respectively. For low relaxation levels the zero-order peaks from a symmetrical and asymmetrical reflections of superlattice can be used to determine an average alloy composition when the relaxation degree has been determined [25]. Therefore we compared our results with lattice parameters for AlxGa1-xN ternary alloy. The lattice parameters shown in Table 1 correspond to AlxGa1-xN compositions x = 0.56, x = 0.54 and x = 0.23, for Sample 1, Sample 2 and Sample 3, respectively. This result agrees well with an average $Al_xGa_{1-x}N$ alloy composition that can be calculated from AlN and GaN thicknesses determined by transmission electron microscopy (see Table 1).

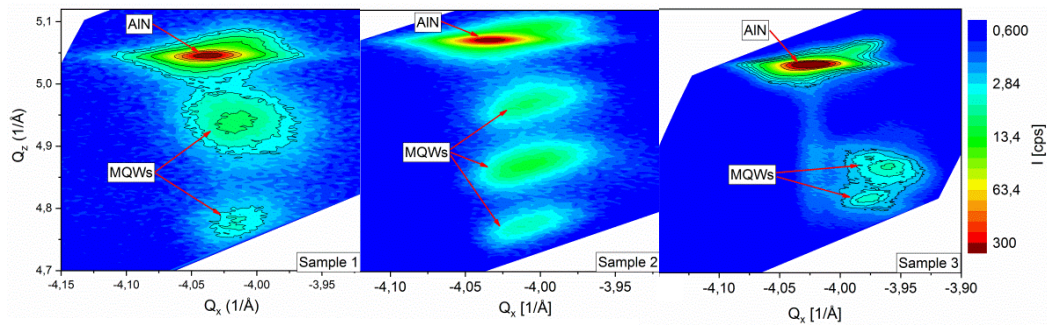

**Figure 3.** Reciprocal space maps of the asymmetrical -1-124 reflection of AlN/GaN MQWs structures. Axes are marked in λ = 2d units (λ = 1.5406 Å - wavelength, d - lattice spacing of (11.4)). Qx axis is in [11.0] direction (parallel to the surface), Qz axis is in [00.1] direction (perpendicular to the surface).

HRXRD and HRTEM results presented in Table 1 show that thicknesses of individual sublayers and values of average lattice parameters of MQWs grown by PAMBE on AlN/sapphire templates are as designed only for Sample 1 and Sample 2. For Sample 3, the AlN barrier grew thinner (2.3 nm) while the GaN barrier grew thicker (6.3 nm) than the expected values of 4 nm. XRD analysis of this sample shows larger degree of lattice mismatch relaxation than for the rest of the samples. However, for all of the samples we observe the tensile in-plane strain in MQWs and compressive strain in the out-of-plane direction. Similar behavior has been observed in the literature [29], [30], [31].

HRXRD maps were used to measure the accurate lattice parameters of MQWs and then to calculate the strain values in our MQWs. For that values of the bulk lattice parameters of the GaN and AlN ($a_{GaN}$ = 3.1893 Å; $a_{AlN}$ = 3.1130 Å; $c_{GaN}$ = 5.1851 Å; $c_{AlN}$ = 4.9816 Å) were used [32, 25]. The results are summarized in Figure 4.

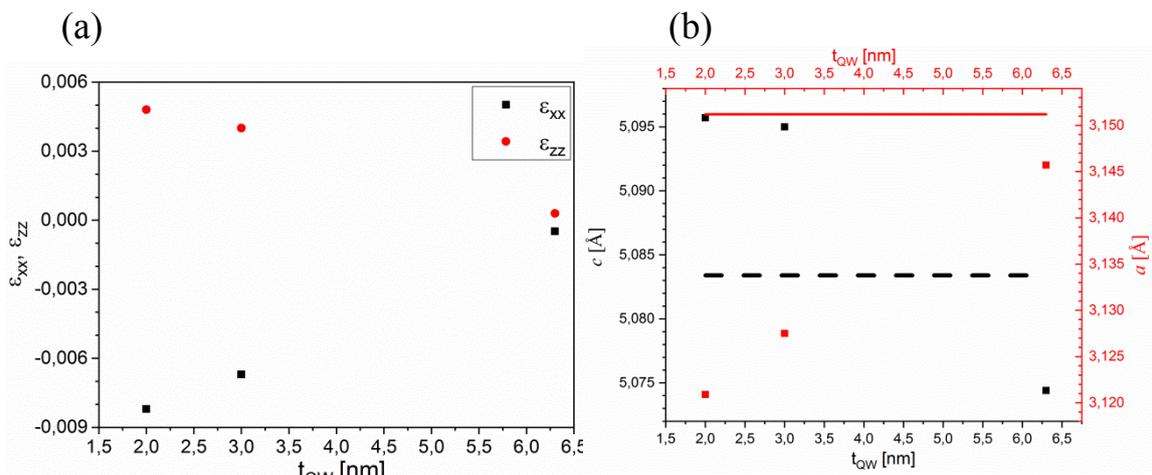

**Figure 4.** $\varepsilon_{xx}$ and $\varepsilon_{zz}$ strain values calculated from HRXRD maps (a) and changes of average lattice constants of the AlxGa1-xN alloy with quantum-well thicknesses (b). To show the trending, the black spotted and red solid lines correspond to $Al_{0.5}Ga_{0.5}N$ lattice parameters $c = 5.0834$ Å and $a = 3.1512$ Å, respectively.

Figure 4 shows the in-plane ($\varepsilon_{xx}$) and out-of-plane ($\varepsilon_{zz}$) strain values as a function of the quantum well thickness. These values are the largest for Sample 1 in which the wells and barriers are the thinnest. With the increase of well/barrier thicknesses the absolute values of out-of-plane (in-plane) strain decrease (increase) and thicker MQWs trend to a relaxed state. For Sample 3 with the thickest wells/barriers we observe the change of the *c*-lattice parameter to the lower value than that of $Al_{0.5}Ga_{0.5}N$. Consequently, for this sample the average composition of Al is smaller than expected for fully strained MQWs with equal thicknesses of GaN wells and AlN barriers. Moreover, this behavior is connected with a higher degree of strain relaxation of MQW which is visible on HRXRD map (Figure 3, Sample 3). A similar observation was reported by Kuchuk et al. [31] for GaN/AlN MQWs and also in different types of wurtzite structures [32].

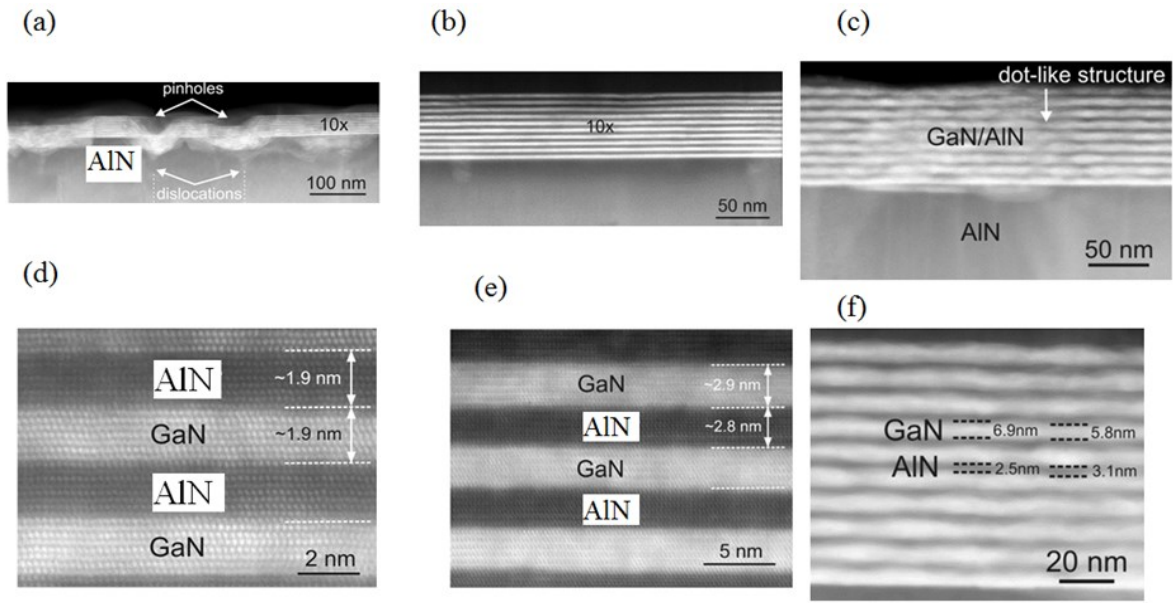

**Figure 5.** Cross-sectional TEM images (upper row) and HRTEM images (lower row) of AlN/GaN MQWs for Sample 1 (a, d), Sample 2 (b, e) and Sample 3 (c, f).

In order to clarify an origin of effects presented above TEM measurements of our MQW structures has been performed. TEM images of the samples are presented in Figure 5. They show thicknesses of GaN and AlN sublayers as designed and a high quality of GaN/AlN interfaces. For Sample 3 some disturbances appear on the interfaces of sublayers. Moreover, the dot-like structure is visible for Sample 3. The results obtained with the TEM technique confirm those from the X-ray studies.

The PL spectra of the investigated samples are presented in Figure 6. Quantum Confined Stark Effect (QCSE) is observed as a red-shift of PL spectra with the increasing well/barrier thicknesses.

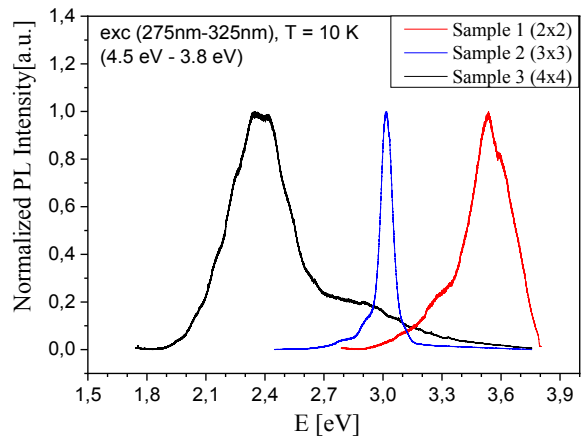

**Figure 6.** Normalized PL spectra measured at low temperature with excitation energy in the range of 4.5 eV to 3.8 eV.

The transition energies (EPL) of the MQW systems were simulated by ab initio calculations using the VASP package. The value of the a lattice parameter was taken from XRD results, and the structure was allowed to relax freely along the [0001] direction to minimize the elastic energy. The details of the calculation method are described in the Ref. [33].

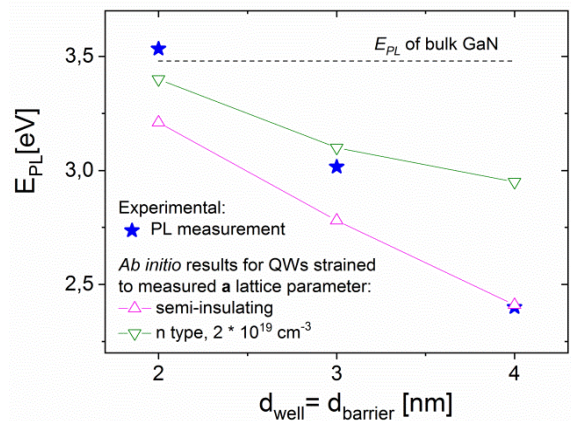

**Figure 7.** PL peak energies as a function of MQW widths, determined experimentally (full stars) and obtained using DFT calculations (open triangles; solid lines are given as guides to the eyes) for semi-insulating and n-doped structures with the *a*-lattice parameters measured by XRD.

The comparison of transition energies determined experimentally and calculated by DFT method is presented in Figure 7. To address the presence of Si intentional dopant, as well as possible unintentional hydrogen and oxygen n-type admixtures which could cause the screening of the built-in electric, theoretical calculations have been performed for two cases: the semi-insulating QWs, and the n-doped QWs with the same level of charge density as in the AlN cap layer. These results clearly show a significant influence of both piezoelectric and screening effects on the emission energy of nitride-based QWs, increasing with the increase of QW width.

The agreement between experimental and theoretical results is quite good. For thick MQWs the experimentally measured value of transition energy is lower than this obtained by theoretical calculations which is most probably caused by the fact that the real thickness of obtained QWs was larger than 4 nm: both XRD and HRTEM results indicated that it exceeded 6 nm. The transition energy observed for 2/2 nm MQWs is slightly higher than the calculated value. There are several possible reasons of this discrepancy. Despite the generally good quality of the structures, thickness fluctuations and dislocations have been observed in TEM images showing the average QW thickness equal to 1.9 nm, a little smaller than 2 nm. Additionally, these defects cause PL peak broadening.

## 4. CONCLUSIONS

Three types of samples with different thicknesses of well/barrier layers were investigated. The increase of SLs thickness leads to strain relaxation of SLs sublayers. We observed that the growth of thin high-quality epitaxial layers is limited by the internal stresses that eventually lead to plastic deformation if they cannot be relaxed by elastic distortion. The more relaxed MQWs were obtained for the thicker well/barrier. TEM analysis confirmed the XRD results. For sample 1 and 3 the existence of pinholes or dot-like structures are registered. It leads to broadening of X-ray diffraction signal and PL spectra. The most efficient PL was detected for sample 2 without observable defects. Moreover, PL signal for sample 3 contains two peaks. It could be the result of defects or the dot-like structure observed in this sample. The red shift of PL spectra is registered with increasing of well/barrier thicknesses, caused by QCSE.

Calculations executed by DFT method are in good agreement with experimental results for the thicker sample. The experimentally measured value of transition energy for thin SL is slightly higher than this obtained by theoretical calculations. The influence of Si intentional dopant in AlN under SLs and in the AlN cap layer as well as possible unintentional hydrogen and oxygen n-type admixtures can cause the screening of the built-in electric field resulting in the increase of transition energy.

## 5. ACKNOWLEDGMENTS

The authors acknowledge support of the Polish National Science Centre grants 2021/43/D/ST7/01936 and 2022/04/Y/ST7/00043 (Weave-Unisono). This research was carried out with the support of the Interdisciplinary Centre for Mathematical and Computational Modelling at the University of Warsaw (ICM UW) under grant no GB84-23.

**CRediT author statement**

**Aleksandra Wierzbicka:** Conceptualization, Methodology, Software, Validation, Formal analysis, Investigation, Writing – Original Draft, Writing – Review & Editing, Visualization, Supervision, Project administration; **Agata Kaminska:** Formal analysis, Visualization, Investigation, Writing – Original Draft, Writing – Review & Editing; **Kamil Sobczak:** Formal analysis, Investigation, Writing – Review & Editing; **Dawid Jankowski:** Formal analysis, Investigation, Writing – Review & Editing; **Kamil Koronski:** Formal analysis, Investigation, Writing – Review & Editing; **Pawel Strak:** Formal analysis, Investigation, Software, Funding acquisition, Writing – Review & Editing; **Marta Sobanska:** Resources, Funding acquisition, Writing – Review & Editing; **Zbigniew R. Zytkiewicz:** Methodology, Resources, Writing – Original Draft, Writing – Review & Editing, Funding acquisition.